\documentclass[twocolumn]{article}
\usepackage{psfig,exscale}

\newcommand{\mbb}[1]{\mbox{\boldmath $#1$}}

\begin{document}

\title{Entanglement transformation at absorbing and amplifying
four-port devices}
\author{S.~Scheel$^1$, L.~Kn\"oll$^1$, T.~Opatrn\'{y}$^{1,2}$, and
D.-G.~Welsch$^1$\\
\small {}$^1$Theoretisch-Physikalisches Institut,
Friedrich-Schiller-Universit\"at Jena, Max-Wien-Platz 1, 07743 Jena,
Germany\\
\small {}$^2$ Department of Theoretical Physics, Palack\'{y} University,
Svobody 26, 771~26 Olomouc, Czech Republic}

\date{\today}
\maketitle

\begin{abstract}
Quantum communication schemes widely use dielectric four-port devices
as basic elements for constructing optical quantum channels. Since for
causality reasons the permittivity is necessarily a complex function of
frequency, dielectrics are typical examples of noisy quantum channels in
which quantum coherence will not be preserved.
Basing on quantization of the phenomenological electrodynamics,
we construct the transformation
relating the output quantum state to the input quantum state without
placing frequency restrictions. Knowledge of the full transformed quantum
state enables us to compute the entanglement contained in the output
quantum state. We apply the formalism to some typical examples in quantum
communication.
\end{abstract}


\section{\hspace*{-1.5ex}Introduction}

Quantum communication experiments widely use dielectric four-port
devices such as beam splitters or optical fibers as basic elements for 
constructing optical quantum channels. Since any frequency-dependent
dielectric function describing an optical element, by virtue of the
Kramers-Kronig relations, is necessarily a complex function of
frequency, absorption is always present which leads to well-known
phenomena as decoherence and entanglement degradation. In order to
study the problem, quantization of the electromagnetic field in the
presence of dielectric media is needed. A consistent formalism
of quantum electrodynamics in absorbing media is reviewed in
\cite{Buch}. It is based on the Green function expansion of the
electromagnetic field with respect to the fundamental variables
of the system composed of the field, the dielectric matter and the
reservoir. All relevant information about the dielectric and
geometric properties are contained in the classical Green function 
of the corresponding scattering problem.

The formalism is especially suited for deriving input-output relations 
of the field at dielectric slabs \cite{Gruner96} on the basis of
measurable quantities as transmission and absorption
coefficients. From the input-output relations we can then derive
closed formulas for calculating the output quantum state from the
(known) input quantum state \cite{Knoll99}. That is, the complete
density matrix after the transformation is known, which makes the
theory most suitable for studying entanglement properties of quantum
states of light. The theory has also been extended to cover amplifying 
media.

In this article we will proceed as follows. The quan\-tum-state
transformation at dielectric four-port devices is shortly
reviewed in Sec.~\ref{state}. An application to entanglement
degradation of Bell states as well as the derivation of separability
criteria for the two-mode squeezed vacuum state are given in 
Sec.~\ref{trans} followed by a summary in Sec.~\ref{conclusions}.

\section{\hspace*{-1.5ex}Quantum-state transformation}
\label{state}

We shortly review the basic formulas needed for the following
considerations. Suppose the electromagne\-tic field has already been
quantized and the Green function for a four-port device
has been rewritten in the form of transmission and absorption
matrices ${\bf T}(\omega)$ and ${\bf A}(\omega)$ \cite{Gruner96}. The
amplitude operators of the incoming and outgoing damped waves at
frequency $\omega$, $\hat{a}_i(\omega)$ and $\hat{b}_i(\omega)$,
respectively, can then be connected by the quantum-optical input-output
relations in the following way:
\begin{equation}
\label{1.1}
{\hat{b}_1(\omega) \choose \hat{b}_2(\omega)}
=
{\bf T}(\omega)
{\hat{a}_1(\omega) \choose \hat{a}_2(\omega)} 
+
{\bf A}(\omega)
{\hat{d}_1(\omega) \choose \hat{d}_2(\omega)} 
\end{equation}
with
\begin{equation}
\label{1.2}
{\bf T}(\omega) {\bf T}^+(\omega)
+\sigma {\bf A}(\omega) {\bf A}^+(\omega) = {\bf I},
\end{equation}
where $\hat{d}_i(\omega)$ represent either device annihilation
operators $\hat{g}(\omega)$ for absorbing devices ($\sigma\!=\!+1$)
or creation operators $\hat{g}^\dagger(\omega)$ for amplifying devices
($\sigma\!=\!-1$). Equation~(\ref{1.2}) is nothing but current (or energy)
conservation. The formulas are valid for any chosen frequency.

By defining the ``four-vector'' operators
\begin{equation}
\label{1.3}
\hat{\mbb{\alpha}}(\omega) =
{\hat{\bf a}(\omega) \choose \hat{\bf d}(\omega)} \,, \quad
\hat{\mbb{\beta}}(\omega) =
{\hat{\bf b}(\omega) \choose \hat{\bf f}(\omega)} \,,
\end{equation}
where $\hat{\bf f}(\omega)$ $\!=$ $\!\hat{\bf h}(\omega)$ for an absorbing
device, and \mbox{$\hat{\bf f}(\omega)$ $\!=$
$\!\hat{\bf h}^\dagger(\omega)$} for
an amplifying device, with $\hat{\bf h}(\omega)$ being some auxiliary
bosonic (``two-vector'') operator. The input-output relation (\ref{1.2})
can then be extended to the four-dimensional transformation
\begin{equation}
\label{1.4}
\hat{\mbb{\beta}}(\omega) = \mbb{\Lambda}(\omega)
\hat{\mbb{\alpha}}(\omega)
\end{equation}
with
\begin{equation}
\label{1.5}
\mbb{\Lambda}(\omega) {\bf J} \mbb{\Lambda}^+(\omega) = {\bf J} \,,
\qquad
{\bf J} =
\left(
\begin{array}{cc}
{\bf I} & {\bf 0} \\ {\bf 0} & \sigma{\bf I}
\end{array}
\right) \,.
\end{equation}
The matrix $\mbb{\Lambda}(\omega)$ is either an element of the compact 
group SU(4) (for absorbing devices) or of the non-compact group
SU(2,2) (for amplifying devices). By introducing the (commuting)
positive Hermitian matrices
\begin{equation}
\label{1.6}
{\bf C}(\omega) \!=\! \sqrt{{\bf T}(\omega) {\bf T}^+(\omega)} , \quad
{\bf S}(\omega) \!=\! \sqrt{{\bf A}(\omega) {\bf A}^+(\omega)} ,
\end{equation}
which, by Eq.~(\ref{1.2}), obey the relation
${\bf C}^2(\omega)\!+\!\sigma {\bf S}^2(\omega)\!=\!{\bf I}$, the
matrix $\mbb{\Lambda}(\omega)$ can be written in the form \cite{Knoll99}
\begin{eqnarray}
\label{1.7}
\lefteqn{
\mbb{\Lambda}(\omega) =
} \nonumber \\[.5ex] && \hspace{-4ex}
\left(\hspace{-1.5ex}
\begin{array}{cc}
{\bf T}(\omega) & {\bf A}(\omega) \\ -\sigma {\bf S}(\omega)
{\bf C}^{-1}(\omega) {\bf T}(\omega) & {\bf C}(\omega)
{\bf S}^{-1}(\omega) {\bf A}(\omega) 
\end{array}
\hspace{-1.5ex}\right)\!.
\end{eqnarray}
It can then be shown that the density operator of the quantum 
state of the outgoing field is given by
\begin{eqnarray}
\label{1.8}
\lefteqn{
\hat{\varrho}_{\rm out}^{({\rm F})} = }
\nonumber \\ && \hspace{-3ex}
{\rm Tr}^{({\rm D})} \big\{ \hat{\varrho}_{\rm in}
\big[ {\bf J} \mbb{\Lambda}^+(\omega) {\bf J}
\hat{\mbb{\alpha}}(\omega), {\bf J} \mbb{\Lambda}^{\!\rm T}(\omega) {\bf J}
\hat{\mbb{\alpha}}^\dagger(\omega) \big] \big\},
\end{eqnarray}
where ${\rm Tr}^{({\rm D})}$ means trace with respect to the
device. Equivalently, the Wigner function (for arbitrary 
$s$-parametrized phase-space functions, see \cite{Scheel00})
transforms as
\begin{equation}
\label{1.9}
W_{\rm out}\big[\mbb{\alpha}(\omega)\big] =
W_{\rm in}\big[{\bf J} \mbb{\Lambda}^+(\omega) {\bf J}
\mbb{\alpha}(\omega)\big] .
\end{equation}

\section{\hspace*{-1.5ex}Entanglement transformation}
\label{trans}

Let us make first a general remark on entanglement transformation. One 
of the requirements to be satisfied by any entanglement measure $E$ is
that a CP map does not increase entanglement
\cite{Vedral98}, that is,
\begin{equation}
\label{2.1}
E\Big(\sum_i \hat{V}_i \hat{\varrho} \hat{V}_i^\dagger\Big) \le
E(\hat{\varrho}), \quad \sum_i \hat{V}_i^\dagger \hat{V}_i = \hat{I}.
\end{equation}
Looking at Eq.~(\ref{1.8}), we see that a quantum-state transformation 
is in fact a CP map for both absorbing and amplifying four-port
devices because in both cases an ancilla (the device) is coupled to
the Hilbert space of the field, a unitary transformation is performed
in the product space
${\cal H}_{\rm field}\otimes{\cal H}_{\rm device}$, and the trace with 
respect to the device is taken at the end. An obvious consequence is
that amplification does not help to increase entanglement.

\begin{figure}[h]
\psfig{file=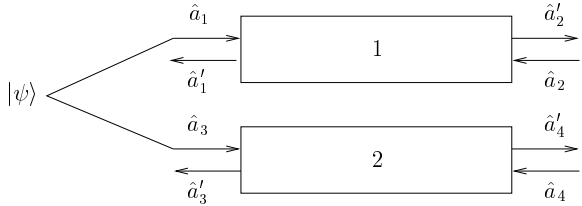,width=7.5cm}
\caption{\label{eightport} A two-mode input field in the
state $|\psi\rangle$ is transmitted through two absorbing
dielectric four-port devices, $\hat{a}_1$, $\hat{a}_3$ ($\hat{a}_2'$,
$\hat{a}_4'$) being the photonic operators of the relevant input
(output) modes.}
\end{figure}

Now we turn to the problem of transmitting two light beams
prepared in an entangled quantum state through absorbing
optical fibers represented by their transmission
coefficients $T_i$ (Fig.~\ref{eightport}).
As an example let us consider the two types of Bell states
\begin{eqnarray}
\label{2.2}
|\Psi^\pm\rangle
&\hspace{-1ex}=&\hspace{-1ex}
\frac{1}{\sqrt{2}}
\big( |01 \rangle \pm |10 \rangle \big) ,\\
|\Phi^\pm\rangle
&\hspace{-1ex}=&\hspace{-1ex}
\frac{1}{\sqrt{2}}
\big( |00 \rangle \pm |11 \rangle \big) .
\end{eqnarray}
Applying Eq.~(\ref{1.8}), after some algebra we get \cite{Scheel00}
\begin{eqnarray}
\label{2.3}
\lefteqn{
\hat{\varrho}_{\rm out}^{({\rm F})}\big( |\Psi^\pm\rangle \big) =
\textstyle\frac{1}{2} \big[ \big( 2\!-\!|T_1|^2\!-\!|T_2|^2 \big)
|00\rangle\langle 00| \big]} \nonumber \\ &&  \hspace{-3ex}
+\textstyle\frac{1}{2} \big( T_2 |01\rangle \pm T_1 |10\rangle \big)
\big( T_2^\ast \langle 01| \pm T_1^\ast \langle 10| \big) ,
\end{eqnarray}
\begin{eqnarray}
\label{2.4}
\lefteqn{
\hat{\varrho}_{\rm out}^{({\rm F})}\big( |\Phi^\pm\rangle \big) =
\textstyle\frac{1}{2} \big[ \big( 1\!-\!|T_1|^2 \big) \big( 1\!-\!|T_2|^2 \big)
|00\rangle\langle 00|
} \nonumber \\ && \hspace{-4ex}
+|T_1|^2 \big( 1\!-\!|T_2|^2 \big) |10\rangle\langle 10|
\!+\!|T_2|^2 \big( 1\!-\!|T_1|^2 \big) |01\rangle\langle 01| \big] 
\nonumber \\ &&  \hspace{-4ex}
+\textstyle\frac{1}{2}
\big( |00\rangle \pm T_1T_2 |11\rangle \big)
\big( \langle 00| \pm (T_1T_2)^\ast \langle 11| \big) .
\end{eqnarray}
The density matrices in Eqs.~(\ref{2.3}) and (\ref{2.4}) have already
been written as a sum of separable states and a single pure state such 
that the convexity property (see, e.g., \cite{Wehrl78})
\begin{equation}
\label{2.5}
E[\lambda \hat{\varrho}_1+(1-\lambda) \hat{\varrho}_2] \le \lambda
E(\hat{\varrho}_1) +(1-\lambda) E(\hat{\varrho}_2)
\end{equation}
can readily be used. Since the entanglement of
a pure state is given by its reduced von Neumann entropy,
the inequality (\ref{2.5}) reduces for equal fibers to
\begin{equation}
\label{2.6}
E\big[\hat{\varrho}_{\rm out}^{({\rm F})}\big( |\Psi^\pm\rangle \big)\big] 
\le |T|^2 \ln 2 ,
\end{equation}
\begin{eqnarray}
\label{2.7}
\lefteqn{
E\big[\hat{\varrho}_{\rm out}^{({\rm F})}\big( |\Phi^\pm\rangle \big)\big] 
} \nonumber \\ && \hspace{-2ex}
\le\textstyle\frac{1}{2} \big[
\big( \!1+\!|T|^2 \big) \ln \big( 1\!+\!|T|^2 \big)
\!-\! |T|^2 \ln |T|^2 \big] .
\end{eqnarray}

\begin{figure}[h]
\psfig{file=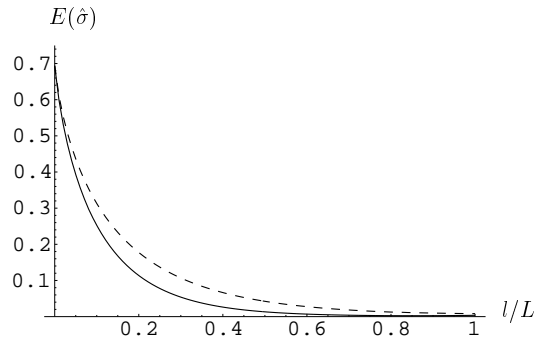,width=7.5cm}
\caption{\label{vergleich} Comparison of entanglement degradation of
one-photon Bell basis states $|\Phi^\pm\rangle$ (full curve) and
$|\Psi^\pm\rangle$ (dashed curve).}
\end{figure}

The numerically calculated relative entropies of the output
quantum states (\ref{2.3}) and (\ref{2.4}) are shown in
Fig.~\ref{vergleich} for equal transmission coefficients of the fibers 
satisfying the Lambert-Beer law
$T\!=\!e^{in\omega l/c}\!=\!e^{in_{\rm R}\omega l/c}e^{-l/L}$ with $l$ 
and $L\!=\!c/(n_{\rm I}\omega)$ being, respectively, the propagation
length through the fibers and the absorption length of the fibers with 
complex refractive index $n(\omega)$ $\!=$ $\!n_{\rm R}(\omega)$
$\!+$ $\!in_{\rm I}(\omega)$.
It is seen that the entan\-gle\-ment of the states
$\hat{\varrho}_{\rm out}^{({\rm F})}\big( |\Phi^\pm\rangle \big)$
decays considerably faster than that of the states
$\hat{\varrho}_{\rm out}^{({\rm F})}\big( |\Psi^\pm\rangle \big)$,
which can be understood from the argument that
in the former case absorption acts on both photons simultaneously.
Since the states considered here live in Hilbert spaces with
dimension 2$\times$2 it would also be possible to compute the exact
amount of entanglement by applying the theorem of Lewenstein and
Sanpera \cite{Lewenstein} which states that in that case the equality
sign in (\ref{2.5}) is realized for a decomposition of the density
matrix in the form
$\lambda\hat{\varrho}_{\rm sep}\!+$
$\!(1-\lambda)\hat{\varrho}_{\rm pure}$ with maximal $\lambda$.

As a second example, let us consider the two-mode squeezed vacuum
state (TMSV) 
\begin{eqnarray}
\label{2.8}
|{\rm TMSV}\rangle
&\hspace{-1ex}=&\hspace{-1ex}
\exp [\zeta (\hat{a}_1^\dagger \hat{a}_2^\dagger
-\hat{a}_1 \hat{a}_2)] |00\rangle \nonumber \\ 
&\hspace{-1ex}=&\hspace{-1ex}
\sqrt{1-q^2} \, \sum\limits_{n=0}^\infty q^n |nn\rangle 
\end{eqnarray}
[$q$ $\!=$ $\!\tanh\zeta$],
which can be used in continuous-variable quantum teleportation
\cite{Braunstein98}. Being difficult to follow the lines presented
above for the Bell states, it is instructive to apply the
separability criterion in Refs.~\cite{Duan00,Simon00} to this
class of states. Since the criterion is based on properties of
the Wigner function for Gaussian states, we can readily use
Eq.~(\ref{1.9}) to obtain a
relation for the bound between separability and
inseparability. Assuming again equal optical fibers with transmission
and reflection coefficients $T$ and $R$ and some thermal photon number
$n_{\rm th}$, we obtain \cite{Scheel00b}
\begin{equation}
\label{2.9}
n_{\rm th} \ge
\frac{(1\!-\!\sigma)(1\!-\!|R|^2)\!+\!|T|^2
(\sigma\!-\!\exp [-2|\zeta|])}{2\sigma
(1-|R|^2-|T|^2)} \,.
\end{equation}
Specifically in the absorbing case ($\sigma$ $\!=$ $\!+1$) we arrive
at the formula for the maximal fiber length $l_{\rm max}$
after which the TMSV
is still nonseparable \cite{Duan00}
\begin{equation}
\label{2.10}
\frac{l_{\rm max}}{L} = \frac{1}{2} \ln \left[ 1+
\frac{1}{2n_{\rm th}} \big( 1-\exp [-2|\zeta|] \big) \right],
\end{equation}
where we have set $R$ $\!=$ $\!0$ and, 
according to the Lambert-Beer law,
$|T|$ $\!=$ $\exp[-l/L]$, with $L$ being again the 
absorption length. On the other hand, from Eq.~(\ref{2.9})
it follows that for an amplifier in the low-temperature limit
($n_{\rm th}$ $\!=$ $\!0$) the maximal square of the absolute
value of the transmission coefficient, $|T_{\rm max}|^2$,
which corresponds to the maximal gain \mbox{$g_{\rm max}$ $\!=$ $\!|T_{\rm
max}|^2$ $\!-$ $\!1$}, is given by 
\begin{equation}
\label{2.11}
|T_{\rm max}|^2 = \frac{2(1-|R|^2)}{1+\exp [-2|\zeta|]} \,,
\end{equation}
which for $R$ $\!=$ $\!0$ reduces to
\begin{equation}
\label{2.12}
|T_{\rm max}|^2 -1 = \tanh|\zeta| = |q|.
\end{equation}
It essentially says that an amplifier that doubles the intensity
of a signal destroys any initially given entanglement.

\section{\hspace*{-1.5ex}Conclusions}
\label{conclusions}

We have shown how the quantum-optical input-output relations at
absorbing and amplifying four-port devices, which follow from
quantum electrodynamics in linear, causal media, can easily be
used to derive bounds on inseparability lengths for optical fields
prepared in Gaussian states and transmitted through noisy quantum
channels such as absorbing or amplifying fibers. For 
finite dimensional Hilbert spaces the quantum-state transformation can 
be used to derive upper bounds on the entanglement content of a given
quantum state of the incoming field. An obvious statement is that
amplifying four-port devices are not able to increase entanglement
since they represent a CP map as in the case of absorbing devices.


\end{document}